%% file: main.tex
\documentclass{article}

\usepackage{arxiv}

\usepackage[utf8]{inputenc} 
\usepackage[T1]{fontenc}    
\usepackage[colorlinks=true, citecolor=black, linkcolor=black, urlcolor=black, filecolor=black]{hyperref}
\usepackage{url}            
\usepackage{booktabs}       
\usepackage{amsfonts}       
\usepackage{nicefrac}       
\usepackage{microtype}      
\usepackage{lipsum}		
\usepackage{graphicx}
\usepackage[numbers]{natbib}
\usepackage{doi}
\usepackage{mathtools}
\usepackage{xcolor}
\definecolor{OliveGreen}{rgb}{0,0.6,0}
\definecolor{CornellRed}{rgb}{0.7, 0.11, 0.11}
\usepackage{caption}
\usepackage{listings}
\usepackage{array, booktabs, multirow, makecell}
\newcolumntype{C}[1]{>{\centering\arraybackslash}p{#1}}
\newcolumntype{L}[1]{>{\arraybackslash}p{#1}}
\usepackage{xcolor,colortbl}
\usepackage{listings}\usepackage{pxfonts}
\usepackage{pifont}
\newcommand{\cmark}{\ding{51}}%
\newcommand{\xmark}{\ding{55}}%
\usepackage{framed}
\usepackage{float}

\title{Test Case-Informed Knowledge Tracing for Open-ended Coding Tasks}

\author{ Zhangqi Duan\\
	University of Massachusetts Amherst\\
	\texttt{zduan@cs.umass.edu} \\
	\And
	Nigel Fernandez \\
	University of Massachusetts Amherst\\
	\texttt{nigel@cs.umass.edu} \\
        \And
	Alexander Hicks \\
	Virginia Tech\\
	\texttt{alexhicks@vt.edu} \\
        \And
	Andrew Lan \\
	University of Massachusetts Amherst\\
	\texttt{andrewlan@cs.umass.edu} \\
}

\date{}

\renewcommand{\shorttitle}{\textit{arXiv} Template}

\hypersetup{
pdftitle={TIKTOC},
pdfsubject={q-bio.NC, q-bio.QM},
pdfauthor={David S.~Hippocampus, Elias D.~Striatum},
pdfkeywords={First keyword, Second keyword, More},
}

\begin{document}
\maketitle

\pagestyle{empty}
\begin{abstract}
Open-ended coding tasks, which ask students to construct programs according to certain specifications, are common in computer science education. Student modeling can be challenging since their open-ended nature means that student code can be diverse. Traditional knowledge tracing (KT) models that only analyze response correctness may not fully capture nuances in student knowledge from student code. In this paper, we introduce Test case-Informed Knowledge Tracing for Open-ended Coding (TIKTOC), a framework to simultaneously analyze and predict both open-ended student code and whether the code passes each test case. We augment the existing CodeWorkout dataset with the test cases used for a subset of the open-ended coding questions, and propose a multi-task learning KT method to simultaneously analyze and predict 1) whether a student's code submission passes each test case and 2) the student's open-ended code, using a large language model as the backbone. We quantitatively show that these methods outperform existing KT methods for coding that only use the overall score a code submission receives. We also qualitatively demonstrate how test case information, combined with open-ended code, helps us gain fine-grained insights into student knowledge. 
\end{abstract}


\section{Introduction}
Open-ended coding questions, or similarly program synthesis tasks, which require students to write open-ended code according to natural language instructions \cite{programsynthesis,adish}, are common in computer science (CS) education. Since these questions require students to \emph{construct} an entire solution, i.e., write code, they may enable education researchers to gain deeper insights into student knowledge compared to other question formats such as multiple-choice. In many domains such as math and/or essay writing, researchers have found evidence that students' open-ended responses contain useful information on their knowledge states, e.g., having misconceptions \cite{brown,feldman,smith} or generally lacking sufficient knowledge \cite{anderson}. A particular advantage of open-ended coding questions in CS education is that student code submissions can be automatically graded, by \emph{compiling} them and passing them through a series of \emph{test cases}; this automated evaluation is usually not applicable in other subjects. Compiling student code can identify syntax errors in student-written code \cite{shaka-etal-2024-error}, while analyzing whether they passed each test case may identify conceptual or logical errors \cite{song2019automatic}; both types of error information can be used to provide timely \emph{feedback} to students to help them correct their errors.

\begin{table}
\caption{Example from our test case-augmented CodeWorkout dataset, which shows a student's code submission to an open-ended coding problem, along with whether it passes a subset of associated test cases.}
\label{tab:qualitative_example}
\small
\centering
\begin{tabular}{p{0.50\linewidth}|p{0.22\linewidth}p{0.08\linewidth}|p{0.03\linewidth}}

\toprule

\multicolumn{4}{L{16cm}}{Problem: A sandwich is two pieces of bread with something in between. Write a Java method that takes in a string str and returns the string that is between the first and last appearance of `bread' in str. Return empty string `' if there are not two pieces of bread.}\\

\midrule

\multicolumn{1}{c|}{\multirow{2}{*}{Student Code Submission}} & \multicolumn{2}{c|}{Test Cases} & \multicolumn{1}{c}{\multirow{2}{*}{Pass/Fail}} \\ 
\cmidrule{2-3}
& Input & Output\\

\midrule

\multirow{-1.8}{*}{\input{student_codes/qual_example_student_code}} & `breadjambread' & `jam' & \multicolumn{1}{c}{\textcolor{OliveGreen}{\cmark}}\\

& `xxbreadjambreadyy' & `jam' & \multicolumn{1}{c}{\textcolor{CornellRed}{\xmark}}\\

& `xxbreadbreadjambreadyy' & `breadjam' & \multicolumn{1}{c}{\textcolor{CornellRed}{\xmark}}\\

& `breadbread' & `' & \multicolumn{1}{c}{\textcolor{OliveGreen}{\cmark}}\\

& `' & `' & \multicolumn{1}{c}{\textcolor{OliveGreen}{\cmark}}\\

& `breaxbreadybread' & `y' & \multicolumn{1}{c}{\textcolor{CornellRed}{\xmark}}\\

& `breadbreadbread' & `bread' & \multicolumn{1}{c}{\textcolor{OliveGreen}{\cmark}}\\

\addlinespace[2pt]
\bottomrule
\end{tabular}
\end{table}

However, existing work on \emph{student modeling} for open-ended questions is limited, possibly due to the noisy nature of open-ended student responses and the scarcity of available data. Evaluation of such models can also be challenging, both quantitatively and qualitatively. The most common evaluation metric is binary \emph{correctness} of student-written code, which corresponds to $1$ if a code submission compiles with no syntax error and passes all test cases, and $0$ otherwise~\cite{shi2022code,pkt}. This evaluation metric is standard in student modeling literature, for both the item response theory (IRT) \cite{irtbook} and knowledge tracing (KT) \cite{kt} model families. Along these lines, existing work has attempted to automatically identify errors and misconceptions among students \cite{shi2021toward} or discover knowledge components (KCs) \cite{shi2023kc}. Despite these functionalities, these works are fundamentally limited by the fact that overall code correctness does not reveal fine-grained insights about student knowledge. More detailed information on whether a student code passes each specific test case is likely necessary to pinpoint specific student errors and even misconceptions. 

Recently, building on rapid advances in large language models (LLMs) and their ability to automatically generate code, researchers have started to develop LLM-powered student models that are capable of not only predicting student code correctness but also \emph{generating} possible student code submissions. These models have the potential to provide fine-grained analysis of student code, extract insights on specific coding styles, errors, and misconceptions, and track how they evolve over time. However, evaluating predicted open-ended codes is much more challenging than evaluating predicted student-written code correctness. The open-ended KT (OKT) framework \cite{okt} replaces the binary-valued response prediction model in existing KT methods with an LLM-based code generator, in order to predict student code submission in a line-by-line, token-by-token way. Despite this new capability, evaluation of code prediction accuracy relies on the CodeBLEU metric \cite{codebleu}, which is influenced by the token n-gram overlap between the predicted code and actual student-written code. As a result, this evaluation is better at capturing \emph{surface code semantic similarity} rather than digging deep into student logic behind their code. The student attempt synthesis framework \cite{adish} uses human experts in-the-loop to evaluate code prediction performance, which is not scalable to real-world application scenarios. For both of these methods, passing the generated code and actual student-written code through test cases and comparing the results will result in a more comprehensive evaluation: test case pass/fail helps us to characterize student code \emph{functionality} in addition to surface semantic similarity. 

\subsection{Contributions}
In this work, we extend KT for open-ended coding questions to incorporate test case-level information, attempting to use them to define a more fine-grained KT task for CS education. Our contributions are summarized as follows:
\begin{itemize}
    \item We define a new KT task that analyzes and predicts whether a student passes each test case in their code submission. This task is more challenging for KT methods since it requires them to make numerous predictions simultaneously for each student code submission. 
    \item We augment the CodeWorkout dataset \cite{codeworkout} with the test cases used for a subset of the questions. We have publicly released this augmented dataset to serve as a benchmark on the test case-level KT task for future KT methods. 
    \item We introduce Test case-Informed Knowledge Tracing for Open-ended Coding (TIKTOC)~\footnote{Our code and supplementary test case data can be found at \url{https://github.com/umass-ml4ed/tiktoc}.}, a novel method for the test case-level KT task. TIKTOC uses a multi-task learning setup to jointly maximize i) test case-level student code submission correctness prediction accuracy and ii) student code generation accuracy, effectively combining and balancing the two types of KT methods outlined above. 
    \item We conduct experiments on the CodeWorkout dataset and show that TIKTOC's multi-task learning setup leads to significant improvement in both test case pass/fail prediction accuracy ($15\%$ gain in AUC) and code prediction accuracy ($6\%$ gain in CodeBLEU) over existing state-of-the-art KT methods adapted to this task. 
    \item We qualitatively show that the new KT task and resulting methods may lead to deeper and more fine-grained insights into student knowledge for open-ended coding questions. We discuss potential use cases of TIKTOC in CS education, along with its limitations, and outline several directions for future work. 
\end{itemize}

\section{Related Work}

\subsection{Knowledge Tracing}

KT \cite{kt} is a well-studied task in the student modeling literature. It studies the problem of breaking down student learning into a series of time steps practicing certain KCs and using the correctness of each response step to track student knowledge. Classic Bayesian knowledge tracing methods \cite{ktcomparepardos,yudelson} use latent binary-valued variables to represent whether a student masters a KC or not. With the increase in popularity of neural networks, multiple deep learning-based KT methods were developed. These models have limited interpretability since student knowledge is modeled as hidden states in complex, deep neural networks. Most of these methods use long short-term memory networks \citep{lstm} or variants \cite{dkt,saint+}, with other variants coupling them with memory augmentation \cite{dkvmn}, graph neural networks \cite{gikt}, attention networks \cite{akt,sakt}, or pre-trained word embeddings or LLMs to leverage question text information \cite{okt, eernna}. These methods vary in how they represent questions and student knowledge; see \cite{abdelrahman2023knowledge} for a survey. KT methods have been applied to many different educational domains, including CS \cite{hoq2023sann, codedkt,pkt}. 

\subsection{Program Synthesis in Computer Science Education} 

There is a line of existing work on analyzing student-generated code, most noticeably using the Hour of Code dataset released by Code.org \citep{peaches,adish,peach}, for tasks such as error analysis and automated feedback generation that are meaningful in CS education settings. 
Program synthesis techniques have been applied for CS education to generate (possibly buggy) student code~\cite{infooirt, okt}, generate new problems~\cite{task_synth} with code explanations~\cite{sarsa2022automatic}, generate student-code guided test cases~\cite{kumar2024using}, provide real-time hints~\cite{hints}, guide students with Socratic questioning~\cite{kumar2024improving}, and suggest bug fixes~\cite{koutcheme2023automated}. However, past work has mostly not incorporated test cases to model student learning in programming. TIKTOC is among the first attempts to analyze and predict whether a student’s code submission passes each test case, potentially offering more fine-grained insights into student knowledge. 

\section{Knowledge Tracing at the Test Case Level}

We now formulate the task of KT at the test case level for open-ended coding problems in CS education. 

\subsection{Problem Formulation}
The goal of the classic KT task is to estimate a student's mastery of KCs/skills/concepts from their responses to past problems and use these estimates to predict their future performance. Formally, given the history of a student's past interactions (e.g., a response to a problem), $x_0, \ldots, x_t$, KT aims to predict aspects of their interaction $x_{t+1}$ (e.g., response correctness on the next problem). For open-ended coding problems, we define a student's interaction with a coding problem as $x_t \coloneq (p_t, \{s^i_t\}, c_t, a_t, \{y^i_t\})$, where $p_t$ is the textual statement of the coding task, $\{s^i_t\}$ are test cases developed to test correctness of code for this problem, $c_t$ is the code submitted by the student, $a_t$ is the score the submission receives, and $y^i_t \in \{0,1\}$ is the test case level past/fail indicator. Here, we use $i$ to index the test cases for each problem. Following existing work~\cite{codedkt}, the binary correctness $a_t$ is equal to $1$ if the code submitted by the student passes all test cases and $0$ otherwise, which provides us a measure of the \emph{overall} correctness of a student submission.

In existing KT methods for CS education~\cite{okt,mao2021knowing, codedkt}, test cases have not been extensively involved in both modeling and evaluation, as we discussed above. However, well-designed test cases can be used to anticipate common logical, syntax, and runtime errors among students, thereby naturally providing fine-grained error insights into nuanced aspects of student knowledge. With this motivation, we investigate a new \emph{test case-level KT task}, which we formalize as: Given a history of student interactions $x_0, \ldots, x_t$ on past programming problems, with each interaction containing information $x_t = (p_t, \{s^i_t\}, c_t, a_t, \{y^i_t\})$, predict which test cases will a student pass/fail, i.e., $\{y^i_{t+1}\}$, on their next attempted problem $p_{t+1}$.

\subsection{Dataset Creation}
\label{sec:dataset}

\begin{figure}

\parbox[t]{6cm}{\null
\centering
\small
  \vskip-\abovecaptionskip
  \captionof{table}[t]{Statistics of our test case-augmented CodeWorkout dataset with $3714$ student code submissions.}%
  \label{tab:dataset_stats}
  \vskip\abovecaptionskip
  
\begin{tabular}{p{0.65\linewidth}|p{0.1\linewidth}|p{0.07\linewidth}|p{0.07\linewidth}|p{0.07\linewidth}}

\toprule

\# unique problems & \multicolumn{4}{c}{$17$}\\
\# unique students & \multicolumn{4}{c}{$246$}\\
\# total no. of test cases & \multicolumn{4}{c}{$305$}\\
\# student code submissions & \multicolumn{4}{c}{$3714$}\\

\midrule
Statistic & $\mu$ & $\sigma$ & Min & Max \\
\midrule
\# test cases/problem & $17.9$ & $4.6$ & $8$ & $26$\\
\# lines/submission & $16.7$ & $7.6$ & $5$ & $82$\\
\# tokens/problem & $67.3$ & $23.1$ & $40$ & $123$\\
\# tokens/test case & $9.8$ & $4.7$ & $3$ & $32$\\
\# tokens/submission & $80.8$ & $34.6$ & $12$ & $344$\\
\# submissions/student & $15.1$ & $3.1$ & $4$ & $17$\\
\# submissions/problem & $218.5$ & $11.1$ & $198$ & $233$\\

\bottomrule
\end{tabular}
  
}
\quad \quad \quad
\hspace{1cm}
\parbox[t]{8cm}{\null
\centering
\includegraphics[width=\linewidth]{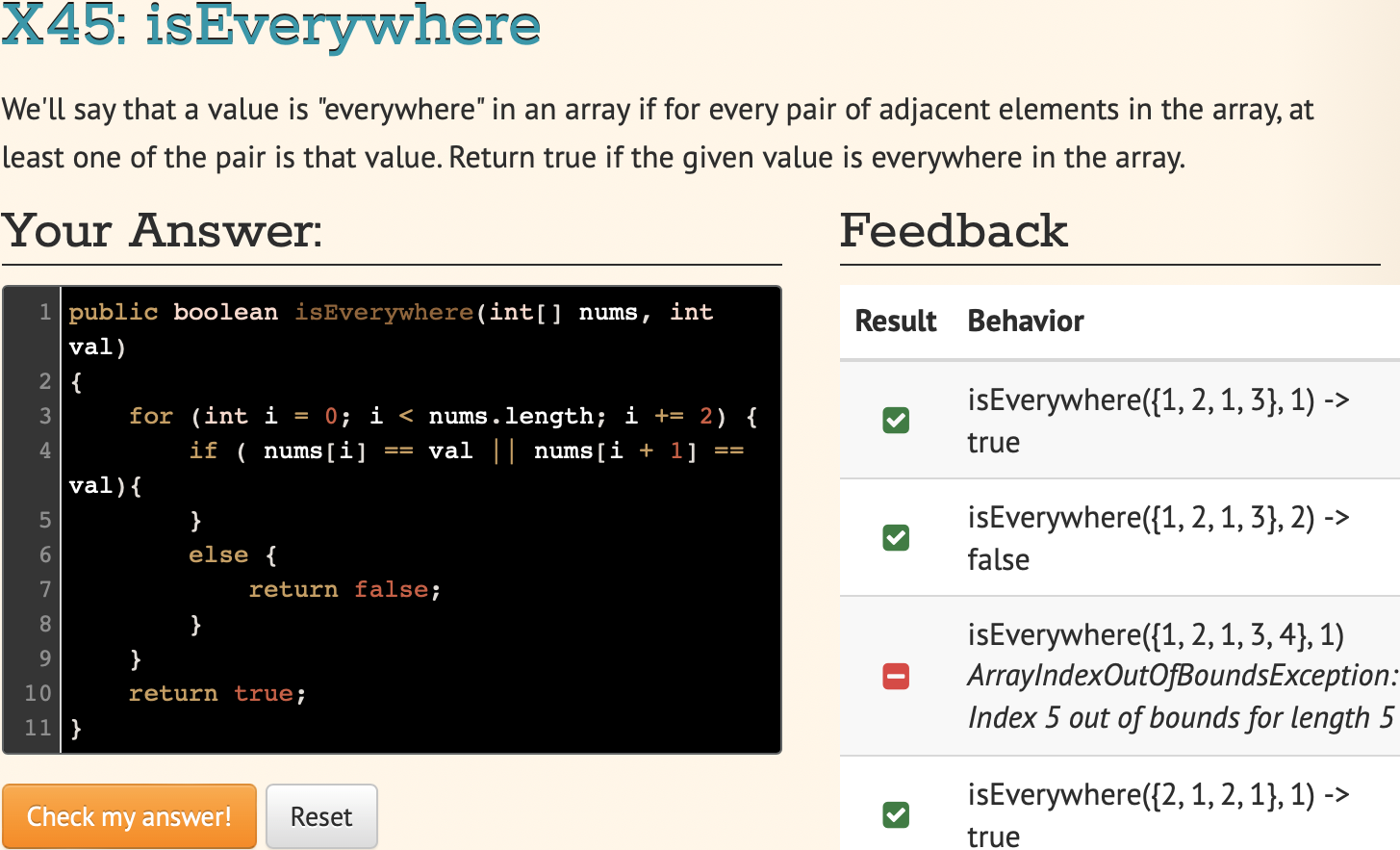}%
\captionof{figure}{The CodeWorkout~\cite{codeworkout} interface with the problem on top, the coding area on the left, and the test case feedback on the right.}%
\label{fig:codeworkout_interface}
}
\end{figure}

We augment the existing CodeWorkout dataset~\cite{codeworkout} with test cases for a subset of the questions. We provide an illustrative example from our test case-augmented CodeWorkout dataset in Table~\ref{tab:qualitative_example}, which includes a sample programming problem, along with a subset of its associated test cases, and a sample code submission made by a student attempting the problem. We provide statistics of the augmented dataset in Table~\ref{tab:dataset_stats} and show the CodeWorkout~\cite{codeworkout} interface in Figure~\ref{fig:codeworkout_interface}. 

The CodeWorkout dataset is a large, real-world programming education dataset previously used in the Second CSEDM Data Challenge~\cite{csedm}. It is a unique, publicly available dataset with university-level Java problems; the dataset contains actual open-ended code submissions from real students, collected from an introductory Java programming course at a large US university. The problems cover various programming concepts including conditionals, and loops, among others. We augment $17$ problems with test cases, with an average of $18$ test cases per problem, leading to $305$ total test cases. These test cases are authentic ones deployed when the dataset was collected: running compilable student code submissions $c_t$ on these test cases recovers the score $a_t$, in the original dataset on PSLC DataShop~\cite{codeworkout}. 

The test cases we extract are a subset of the original test cases used by CodeWorkout during the collection of the CSEDM dataset in 2019~\cite{csedm}. Historically, CodeWorkout has supported three methods of writing instructor-provided tests: JUnit test cases written by exercise authors, output validation tests that provide input and match output from the execution of the exercise to an expected value, and JUnit tests that were interpolated into base JUnit test file server side; we extract test cases of the second type that are written by course staff and CodeWorkout developers. The test cases were designed to evaluate the correctness of the student submission by checking for common logical errors in the students' code. Test cases can be both public and hidden: public ones often seek to remind the students to consider specific edge cases, while hidden ones will reserve one logical fallacy or edge case as an exercise for the student to identify. For example, Table~\ref{tab:qualitative_example} includes a test case that inputs an empty string (`'). This test is an example of an edge case check, verifying the behavior of the student's code at a boundary value that they might not consider as valid input. Another good test case (`breadbreadbread') checks whether students handle multiple instances of the keyword token correctly, by ensuring they properly grab the first and last instances in a string, instead of a common misconception, the first and second instances. There are often multiple variants of easy test cases (`breadjambread') to give students credit and encourage them, while they incrementally build and test their code~\cite{edwardsProgressIndicatorsMeasuring2016}.

We leverage these test cases associated with each problem to obtain \textit{test case-level pass/fail labels}, i.e., $\{y^i_t\}$, for each student code submission. We build an automated evaluation pipeline that first compiles a student code submission for a problem. If a student code fails to compile, we mark all test cases associated with the problem as failed for this submission. To account for cases like infinite loops, we add a timeout condition in our pipeline: If a student code times out after $30$ seconds when attempting compilation, we also mark all associated test cases as failed. If a student code compiles, our pipeline automatically runs the code against all associated test cases. If the output of the code matches the expected output obtained by running the correct solution code, we mark the test case as passed and as failed otherwise. 

\section{Methodology}

KT at test case level is a novel task with \textit{no existing methods}. Therefore, we build on two strong, state-of-the-art KT methods for programming, Code-based Deep KT (Code-DKT)~\cite{codedkt} and Open-Ended KT (OKT)~\cite{okt}, to develop a multi-task learning approach and apply it to the test case level KT task. In this section, we first introduce these methods before detailing our novel method, TIKTOC, for KT at test case level.

\subsection{Code-DKT}
\label{sec:codedkt}

Code-DKT~\cite{codedkt} is a KT method that leverages the content of student code to improve on DKT~\cite{dkt} for open-ended coding tasks. Instead of using only the binary-valued correctness of the last student response as input, Code-DKT uses code2vec~\cite{code2vec}, a neural code representation model, to obtain a meaningful representation of past student code submissions. To match the original DKT's data format, Code-DKT characterizes a student code submission as binary-valued, aggregated across all test cases: if all test cases pass, the submission is correct, otherwise, it is incorrect. Therefore, an interaction is simplified to $x_t \coloneq (p_t, c_t, a_t \in \{0,1\})$. For each student, at each timestep $t$ (i.e., an interaction or student submission), Code-DKT updates the student's estimated knowledge state by 
\begin{align*}
    h_{t} = \text{LSTM}(h_{t-1}, c_t, p_t, a_t),
\end{align*}
through a long short-term memory (LSTM) network \cite{lstm}; the input to the LSTM update module is a combination of the code2vec embedding of the student code, $c_t$, and a one-hot encoded representation of the last problem-response correctness, $(p_t, a_t)$, like DKT~\cite{dkt}. To predict the correctness of the next student submission to the problem $p_{t+1}$, Code-DKT uses a linear prediction head and a sigmoid function:
\begin{align}
    \hat{a}_{t+1} = \sigma(W \cdot h_{t}),
\end{align}
where $W$ denotes the parameters of a learnable linear layer. 

Code-DKT then minimizes the binary cross entropy (BCE) loss for submission correctness prediction from one student code submission:
\begin{equation*}
    \mathcal{L}_{\text{Code-DKT}} = a_t \cdot \log \hat{a}_t + (1-a_t)\cdot \log (1-\hat{a}_t)
\end{equation*}
with the final objective being the mean of this loss over code attempts by all students to all attempted problems. 

Code-DKT and other KT methods predicting overall correctness are fundamentally limited by their lack of fine-grained insights about student knowledge. On the other hand, leveraging test cases and predicting which test cases a student code would pass could provide specific student errors and misconceptions.

\subsection{OKT}
\label{sec:okt}

OKT~\cite{okt} is the first KT method that is generative in nature; it leverages LLMs to predict the open-ended code submitted by a student on their next programming problem in a line-by-line, token-by-token way, instead of just predicting the binary-valued correctness of their code.

In OKT, the input to the KT model relies on a different embedding method to encode past student code submissions than Code-DKT. For student code, OKT first transforms it into an Abstract Syntax Tree (AST), followed by using the ASTNN~\cite{astnn} model to obtain an embedding preserving the syntactic and semantic features of the code. Moreover, as an improvement over Code-DKT, OKT also encodes the textual problem statement using text embedding methods to include as part of the input to the KT model. The KT model in OKT is flexible and can be adapted from any existing binary-valued KT method. For example, if we choose DKT as the KT model, then we simply use the encoded problem statement and student code from the previous response as input to the LSTM model, just like Code-DKT.

The main difference between OKT and all other existing KT methods is that it uses an LLM-based response generation component to predict the next student code submission, $c_{t+1}$, given the current knowledge state, $h_t$, and the statement of the next problem, $p_{t+1}$. Specifically, the next problem text $p_{t+1}$ is tokenized by an open-source LLM into a sequence of $M$ tokens, where each token has a $D$-dimensional embedding, i.e., $\bar{p}_m \in \mathcal{R}^{D}$ for $m = 1, \ldots, M$ (we drop the timestep $t+1$ in denoting problem tokens for simplicity). Then, OKT injects the knowledge state of the student $h_t$ into the LLM, creating \textit{knowledge-guided} token embeddings using a linear alignment function $f$, i.e., $p_m = f(\bar{p}_m, h_t)$. These knowledge-guided problem embeddings are fed into Llama 3 to generate the predicted student code, token-by-token. The loss to be minimized for one student code submission is given by:
\begin{align*}
    \mathcal{L}_{\text{OKT}} = \frac{1}{N}\sum_{n=1}^N -\log P_{\theta}\left( c^n\ \middle \vert\ p_m,\{c^{n'}\}_{n'=1}^{n-1}\right),
\end{align*}
where $N$ is the number of tokens in the student code submission. $\theta$ denotes the set of learnable parameters, which includes the underlying KT model, the alignment function, and the LLM's parameters (if fine-tuned). The final training objective is the mean of this loss over code submissions by all students to all attempted problems. 

OKT analyzes richer information than Code-DKT in the form of open-ended student code. However, comparing predicted code and actual student-written code only captures surface code semantic similarity. We need to look deeper to gain fine-grained insights into student knowledge and identify specific logical errors or misconceptions.

\subsection{TIKTOC}
\label{sec:tiktoc}

\begin{figure}
  \centering
  \includegraphics[width=\linewidth]{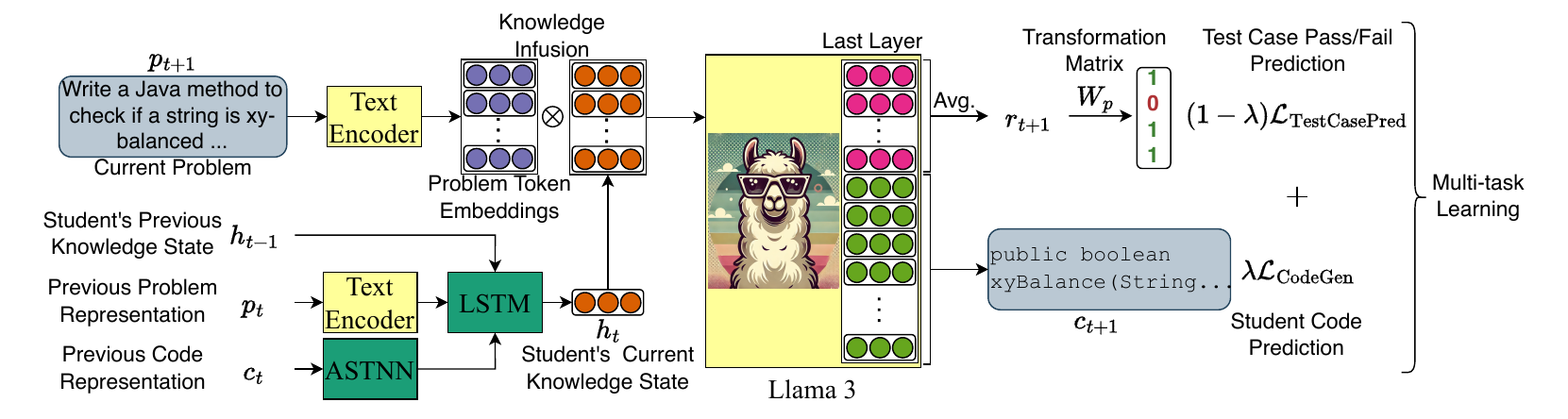}
  \caption{ 
  Overview of TIKTOC's model architecture with the Llama 3 LLM as the backbone. TIKTOC embeds the student's previous open-ended code to update the student's knowledge estimate, which is then combined with the current problem, as input to Llama 3. TIKTOC simultaneously learns to predict both 1) whether a student’s code submission passes each test, and 2) the student’s open-ended code, with a multi-task learning setup.}
  \label{fig:model}
\end{figure}

We now detail our TIKTOC model, illustrated in Figure~\ref{fig:model}. TIKTOC has two key innovations: First, we utilize test cases and model whether a student attempt passes each single test case, potentially offering more fine-grained insights into student knowledge. Second, we use a multi-task learning setup to combine the objectives of both Code-DKT and OKT: TIKTOC simultaneously analyzes and predicts 1) whether a student’s code submission passes each test case, and 2) the student’s open-ended code. Intuitively, these two tasks can be mutually beneficial to each other, which motivates the multi-task learning approach. For example, failing a test case evaluating a loop termination conditional significantly reduces the space of possible code, which enables better student code prediction reflecting this error, and vice versa. 

\paragraph{Setup} TIKTOC combines and builds upon elements from both Code-DKT and OKT into a unified model that uses a single LLM as the backbone. For each student, at each timestep $t$, TIKTOC first converts their sequence of past interactions $x_0, \ldots, x_t$ attempting programming problems up to timestep $t$ into continuous representations, where each interaction $x_t \coloneq (p_t, c_t, \{y^i_t\})$ contains the attempted problem, their code submission, and pass/fail outcomes of the test cases associated with the problem. Similar to OKT~\cite{okt}, for the problem representation, TIKTOC uses the mean problem token embeddings from Llama 3, and for the code representation, uses ASTNN~\cite{astnn} to embed student code.

\paragraph{Code generation} The code generation setup for TIKTOC is similar to OKT~\cite{okt}, with DKT as the underlying KT model, except that we use Llama 3~\cite{llama3}, a more recent, more powerful open-source LLM as the code generation backbone. We use the same way to inject student knowledge states into the LLM embedding space, with one modification: we use hidden knowledge states with $D=4096$ dimensions to align with Llama 3's text embedding space. The loss for one student code submission is given by
\begin{align} \label{eq:cgloss}
    \mathcal{L}_{\text{CodeGen}} = \sum_{n=1}^N -\log P_{\theta}\left( c^n\ \middle \vert\ p,j,\{c^{n'}\}_{n'=1}^{n-1}\right),
\end{align}
with the final $\mathcal{L}_{\text{CodeGen}}$ loss being the mean over code attempts by all students to all attempted problems.

\paragraph{Test case prediction} For the task of predicting individual test case pass/fail outcomes for a student code submission, we modify the setup of Code-DKT's prediction head. We average the hidden states of the last layer of Llama 3 that correspond to only the input, i.e., the knowledge-guided problem statement embeddings, to obtain a representation $r$ that we use for predictions. We choose to not include the output code prediction hidden states in our average, since including them means that the predictions are conditioned on generated code, which can simply be done by the code compiler in a non-probabilistic way. Building on Code-DKT~\cite{codedkt}, we transform $r$ into a prediction vector $\hat{y}$ through a transformation matrix $W_p$, i.e.,
\begin{align*}
    \hat{y} = \sigma(W_{p}r).
\end{align*}
The prediction vector has $|S_p|$ dimensions, where $|S_p|$ is the number of test cases associated with problem $p$. Here, element $i$ of the vector $\hat{y}$ represents the probability of the student submission passing test case $s^i_p$. The BCE loss for test case pass/fail outcome prediction for one student code attempt is given by
\begin{align} \label{eq:tcploss}
    \mathcal{L}_{\text{TestCasePred}} = \frac{1}{|S_p|}\sum_{i=1}^{|S_p|} y^i\cdot \log (\hat{y}^i) + (1-y^i)\cdot \log (1-\hat{y}^i).
\end{align}
with the final $\mathcal{L}_{\text{TestCasePred}}$ loss being the mean over code attempts by all students to all attempted problems. 

We learn a separate transformation matrix $W_p$ for each problem since the number and nature of test cases differ across problems. We experiment with different variants of parameterizing $W_p$ which we detail in Section~\ref{sec:ablation}. We find that treating test cases as one-hot encodings performs best. We randomly initialize $W_p$ and set the number of columns to be equal to the number of test cases associated with problem $p$, with each column thereby learning a representation of its corresponding test case.

\paragraph{Multi-task learning} Our final multi-task objective minimizes a combination of both losses together, i.e., the code generation loss $\mathcal{L}_{\text{CodeGen}}$ in Eq.~\ref{eq:cgloss} and the test case outcome prediction loss $\mathcal{L}_{\text{TestCasePred}}$ in Eq.~\ref{eq:tcploss}, with a balancing parameter $\lambda \in [0,1]$ controlling the importance of the two losses, i.e.,
\begin{align} \label{eq:tiktocloss}
    \mathcal{L}_{\text{TIKTOC}} = \lambda \mathcal{L}_{\text{CodeGen}} + (1-\lambda) \mathcal{L}_{\text{TestCasePred}}.
\end{align}
The optimal value of $\lambda$ can be learned~\cite{kendall2018multi} as a parameter of the model or through a grid search. In our experiments, we find TIKTOC significantly outperforms baseline KT methods with an initial guess of $\lambda=0.5$ making a grid search redundant. We hypothesize that both objectives offer valuable insights into the underlying KT problem and can benefit each other, improving over standalone versions of both Code-DKT and OKT. 

\section{Experiments}
\subsection{Metrics}

For the test case-level KT task, since the pass/fail status of each student code submission on each test case is binary-valued, we evaluate the performance of KT methods using standard metrics such as AUC, F1 Score, and accuracy~\cite{dkt,codedkt}. Since the test case pass/fail outcome labels can be imbalanced, AUC and F1 Score are important.
For the \textit{student code prediction} task, following OKT~\cite{okt}, we evaluate model-generated student codes against ground-truth student codes using CodeBLEU~\cite{codebleu}, a variant of the classic text similarity metric BLEU~\cite{papineni2002bleu}, adapted for code. This metric measures both the syntactic and semantic similarity between two pieces of code. To test whether models simply memorize frequent student code in the training data, we also evaluate predicted \textit{student code diversity}. We measure code diversity by using the popular dist-N metric~\cite{li-etal-2016-diversity} which computes the ratio of unique N-grams to the total number of N-grams. 

\subsection{Baselines}

KT at the test case level is a novel task with \textit{no existing methods}. Therefore, we adapt both Code-DKT~\cite{codedkt} and OKT~\cite{okt} to this task, to serve as strong baselines.

We adapt the standard Code-DKT method, detailed above in Sec.~\ref{sec:codedkt}, to our test case-level KT task, which we refer to as \textbf{Code-DKT-TC}. Specifically, we modify Code-DKT's prediction head to predict individual test case pass/fail outcomes for a student code submission in a similar fashion as TIKTOC. We concatenate the embedding of the next problem $p_{t+1}$ from Llama 3~\cite{llama3} with the student's estimated knowledge state, $h_t$, from the LSTM~\cite{lstm}. We then transform it into a prediction vector, $\hat{y}$, through a non-linear transformation, $\hat{y} = \sigma(W_{p}(p_{t+1} \oplus h_t))$, where $\oplus$ denotes vector concatenation and $W_p$ denotes a problem-specific learable transformation matrix. The training objective for Code-DKT-TC is the same as the test case prediction objective of TIKTOC in Eq.~\ref{eq:tcploss}. To ensure a fair comparison, we estimate a student's knowledge state in the same way as in OKT, using Llama 3's mean token embeddings of the problem statement and the ASTNN embedding of student code as input to the LSTM. We switch from the original code2vec~\cite{code2vec} representation to ASTNN~\cite{astnn} to embed student codes and use ASTNN across methods to align with the settings in OKT. This switch is also motivated by the observation that ASTNN outperforms code2vec in embedding student codes when modeling student learning in programming~\cite{mao2021knowing}.

We also adapt OKT, detailed in Sec.~\ref{sec:okt}, to our test case-level KT task, which we refer to as \textbf{OKT-TC}. This baseline simply takes the predicted student code submission generated by OKT, compiles it, and evaluates it on the test cases. This evaluation pipeline follows recent work~\cite{shaka-etal-2024-error} on detecting errors/bugs present in predicted code by OKT. Same as before, if the predicted code fails to compile or times out during compilation, we mark all associated test cases as failed. We change the base LLM of OKT~\cite{okt} from GPT-2~\cite{gpt2} to the more recent and powerful Llama 3~\cite{llama3}. We use Llama 3 as the base LLM across all KT methods we experiment with in this work to ensure a fair comparison.

As a sanity check, to estimate the difficulty of our test case-level KT task and to estimate a lower bound of performance, we also incorporate two simple baselines: \textbf{Random}, which simply predicts the test case pass/fail outcome of a student code randomly with equal probability, and \textbf{Majority}, which simply predicts the test case pass/fail outcome of a student code as pass since pass occurs more frequently than fail.

\subsection{Experimental Setup}

Following prior work \cite{codedkt}, we experiment on the first code submission for each student on each problem in the CodeWorkout dataset. To ensure a fair comparison across KT methods, we use the instruction-tuned version of Llama 3~\cite{llama3} with $8$B parameters as the base LLM, and a frozen ASTNN~\cite{astnn} code embedding model to embed student codes following OKT~\cite{okt}. We use the Parameter Efficient Fine-Tuning (PEFT) library from HuggingFace~\cite{wolf-etal-2020-transformers} 
to load Llama 3 $8$B Instruct, and train via low-rank adaptation (LoRA)~\cite{hu2022lora} ($\text{LoRA }\alpha=256, \text{LoRA }r=128, \text{LoRA dropout}=0.05$) using $8$-bit quantization~\cite{dettmers2024qlora}. We use the AdamW~\cite{loshchilov2018decoupled} optimizer with a
batch size of $32$. 
We perform a grid search to find the optimum learning rate (LR). For TIKTOC and OKT, we use an LR of $1e-5$ for Llama 3 8B Instruct, $5e-5$ for the LSTM, and $1e-4$ for the transformation matrix $W_p$ as well as the linear alignment function $f$. We use a linear LR scheduler with warmup and perform gradient clipping. 
For Code-DKT-TC, we use an LR of $1.5e-3$ for the LSTM and $1e-3$ for the transformation matrix $W_p$, with the ReduceLROnPlateau LR scheduler.

For Code-DKT-TC and TIKTOC, we learn a separate transformation matrix $W_p$ for each problem, with each column representing an associated test case using $D=4096$ dimensions. For OKT, we use the Java compiler from OpenJDK $11.0.23$ to compile and run predicted student codes against test cases. For TIKTOC, we find that simply setting the multi-task balancing parameter in Eq.~\ref{eq:tiktocloss} to $\lambda=0.5$ works well across all settings. A likely cause is that once normalized over all test cases and all code tokens, both objectives are BCE losses and are already on the same scale, which alleviates the need for a grid search over this parameter. For OKT and TIKTOC, we use greedy decoding to generate student code. We fine-tune for $20$ epochs with early stopping on the validation set on a single NVIDIA L$40$S $48$GB GPU, with each epoch taking up to $1$, $35$, and $45$ minutes, for Code-DKT, OKT, and TIKTOC, respectively.

\section{Results, Analysis, and Discussion}

In this section, we quantitatively evaluate model performance on test case-level KT and student code prediction, perform an ablation study on TIKTOC, and compare TIKTOC with other KT methods through qualitative case studies.

\subsection{Quantitative Results}

\begin{table*}
\caption{Performance on 1) test case-level KT and 2) student code prediction for all approaches across all metrics. TIKTOC, with a multi-task learning setup with the Llama 3 LLM as the backbone, outperforms existing KT approaches by a wide margin. Standard deviation across $5$ random seeds are shown in parentheses. Best performance is in \textbf{bold} and second best is \underline{underlined}.}
\label{tab:results}
\small
\centering
\begin{tabular}{p{0.25\linewidth}p{0.12\linewidth}p{0.12\linewidth}p{0.12\linewidth}|p{0.12\linewidth}p{0.12\linewidth}}

\toprule

\multirow{3}{*}{Model} & \multicolumn{3}{c}{Test Case-level KT Performance} &  \multicolumn{2}{c}{Student Code Prediction}\\
\cmidrule{2-6}
& AUC $\uparrow$ & F1 Score $\uparrow$ & Accuracy $\uparrow$ & CodeBLEU $\uparrow$ & Dist-1 $\uparrow$\\

\midrule

Random & $0.501\ (0.8\%)$ & $0.545\ (2.5\%)$ & $0.502\ (0.6\%)$ & $-$ & $-$\\ 
Majority & $0.501\ (0.3\%)$ & $0.719\ (8.9\%)$ & $0.620\ (5.5\%)$ & $-$ & $-$\\ 
Code-DKT-TC~\cite{codedkt} & $\underline{0.661}\ (4.3\%)$ & $\underline{0.771}\ (2.4\%)$ & $\underline{0.677}\ (2.5\%)$ & $-$ & $-$\\ 
OKT-TC~\cite{okt} & $-$ & $0.763\ (3.3\%)$ & $0.647\ (3.2\%)$ & $\underline{0.522}\ (2.1\%)$ & $\textbf{0.383}\ (1.3\%)$\\ 
TIKTOC (ours) & $\textbf{0.764}\ (4.3\%)$ & $\textbf{0.794}\ (4.5\%)$ & $\textbf{0.723}\ (3.0\%)$ & $\textbf{0.554}\ (1.0\%)$ & $\underline{0.369}\ (5.9\%)$\\ 

\bottomrule
\end{tabular}
\end{table*}

Table~\ref{tab:results} shows the average performance (and standard deviation) on test case pass/fail prediction and code generation, across $5$ random folds, for all methods. We see that the simple Random and Majority baselines perform poorly, which suggests that test case-level student response prediction is inherently difficult. This difficulty comes from the need to simultaneously predict numerous test cases for each student code submission. Code-DKT-TC outperforms OKT-TC on test case level pass/fail prediction, although it cannot perform the code generation task. This result can be explained by Code-DKT-TC being explicitly trained on the test case-level prediction objective; since OKT-TC evaluates test case pass/fail after the entire predicted student code is generated, the discrete nature of code makes this process non-smooth, lowering the robustness of the prediction task.

We see that our proposed approach, TIKTOC, outperforms the next best baseline on test case pass/fail prediction, Code-DKT-TC, by a wide margin of $15.6\%$ on the AUC metric. TIKTOC also outperforms the best baseline on student code prediction, OKT-TC, by $6.1\%$ on the CodeBLEU metric. TIKTOC achieves a statistically significant performance improvement over Code-DKT-TC/OKT-TC, the next best baseline(s) on test case-level KT/student code prediction, with p-values of $0.005$/$0.02$, respectively, obtained via a paired t-test. These results validate our hypothesis that test case-level pass/fail prediction and student code prediction are two different tasks that are beneficial to each other; TIKTOC effectively leverages this symbiosis with a multi-task learning approach.

\subsection{Ablation Study}
\label{sec:ablation} 

Table~\ref{tab:results_ablation} shows the results of an ablation study for TIKTOC. We see that combining student knowledge states with the LLM's input text embeddings is crucial and removing the student model from TIKTOC results in a large drop in performance, as seen in row \textbf{No Knowledge Estimation}. This observation highlights the importance of the underlying student model. 

There can be many ways to represent different test cases in TIKTOC. We find that treating test cases as one-hot encodings performs best, as seen in row \textbf{One-hot Test Case}, where we initialize the embeddings of test cases independently at the start of the model training process. Alternatively, instead of random initialization, we experiment with initializing each test case representation with their mean token embedding from Llama 3~\cite{llama3}. This method, shown in row \textbf{Embed Test Case}, leverages the textual information of the test cases. We also experiment with embedding both the problem content and that of test cases, shown in row \textbf{Embed Test Case with Problem}. However, we see that neither of these approaches improves performance over one-hot encoding test cases, which is counterintuitive; one possible reason is that the test cases are usually functions and not purely textual in nature, which can be difficult for an LLM to capture. Future work needs to develop ways to effectively leverage the content of test cases to improve performance on the test case pass/fail prediction task. 

\begin{table*}
\caption{Ablation study of TIKTOC comparing different model variants. Best performance is in \textbf{bold} and second best is \underline{underlined}.}
\label{tab:results_ablation}
\small
\centering
\begin{tabular}{p{0.33\linewidth}p{0.1\linewidth}p{0.1\linewidth}p{0.1\linewidth}|p{0.11\linewidth}p{0.1\linewidth}}

\toprule

\multirow{3}{*}{TIKTOC Model Ablations} & \multicolumn{3}{c}{Test Case-level KT Performance} &  \multicolumn{2}{c}{Student Code Prediction}\\
\cmidrule{2-6}
& AUC $\uparrow$ & F1 Score $\uparrow$ & Accuracy $\uparrow$ & CodeBLEU $\uparrow$ & Dist-1 $\uparrow$\\

\midrule

No Knowledge Estimation & $0.699$ & $0.799$ & $0.648$ & $0.517$ 
& $0.357$\\ 
Embed Test Case  & $\underline{0.757}$ & $\underline{0.834}$ & $\underline{0.792}$ & $0.532$ 
& $0.360$\\ 
Embed Test Case with Problem & $0.744$ & $0.828$ & $0.786$ & $\underline{0.55}$ 
& $\underline{0.366}$\\ 
One-hot Test Case (Main Model) & $\textbf{0.764}$ & $\textbf{0.837}$ & $\textbf{0.802}$ & $\textbf{0.566}$ 
& $\textbf{0.369}$\\ 

\bottomrule
\end{tabular}
\end{table*}

\subsection{Qualitative Case Studies}

We now qualitatively show that the test case pass/fail prediction and code generation objectives in our multi-task learning setup are mutually beneficial to each other with three qualitative case studies.

\paragraph{Code generation helps test case pass/fail predictions.} 
Table~\ref{tab:qualitative_codegen_helps_testcasepred} shows a sample interaction from the test set, where a student attempts a problem to check whether a string is ``xy-balanced''. TIKTOC accurately predicts a semantically close student code submission to ground-truth student code submission, albeit with different variable names. Code-DKT incorrectly predicts that the two harder test cases, the empty string `', and the string `bbb', without either `x' or `y', which test edge cases in a student's logic, will fail. TIKTOC, on the other hand, can leverage its knowledge of the anticipated student code, to implicitly run this code against its test case representations, thereby imitating a compiler, to accurately predict the test cases would pass.

\paragraph{Test case pass/fail predictions help code generation.}
Table~\ref{tab:qualitative_testcasepred_helps_codegen} shows a sample interaction where a student attempts a problem to check whether three integers are evenly spaced. TIKTOC correctly predicts all test case pass/fail labels matching the ground-truth labels. In particular, test cases containing three unsorted evenly spaced integers, for example, $(4,6,2)$ and $(6,2,4)$, are correctly predicted as failed. OKT, which does not have access to these valuable test case pass/fail predictions, incorrectly predicts a student code that will pass all test cases, using constructs such as ``Math.abs'' not present in the ground-truth student code. TIKTOC, on the other hand, can leverage its knowledge of which test cases are predicted to fail, to predict a semantically closer student code to the ground-truth student code. It successfully predicts a similar bug that reflects the student's logical error in assuming the three input integers are sorted. 

\begin{table}
\caption{Qualitative case study showing code generation helps test case pass/fail predictions. TIKTOC can leverage its knowledge of anticipated student code to implicitly run this code against its test case representations. As a result, it can accurately predict test case pass/fail outcomes. A subset of test cases is shown. The code indentation is changed for brevity.}
\label{tab:qualitative_codegen_helps_testcasepred}
\small
\centering
\renewcommand{\arraystretch}{1.1}
\begin{tabular}{p{0.28\linewidth}|p{0.28\linewidth}|p{0.06\linewidth}p{0.05\linewidth}|p{0.05\linewidth}p{0.07\linewidth}p{0.03\linewidth}}

\toprule

\multicolumn{7}{L{16cm}}{Problem: A string is xy-balanced if for all the `x' characterss in the string, there exists a `y' character somewhere later in the string. So `xxy' is balanced, but `xyx' is not. One `y' can balance multiple `x'. Return true if the given string is xy-balanced.}\\

\midrule

\multicolumn{1}{c|}{\multirow{3}{*}{Ground-truth Student Code}} & \multicolumn{1}{c|}{\multirow{3}{*}{TIKTOC Predicted Student Code}} & \multicolumn{2}{c|}{Test Cases} & \multicolumn{3}{c}{\textcolor{OliveGreen}{\cmark}/\textcolor{CornellRed}{\xmark} Preds} \\ 
\cmidrule{3-7}
&& Input & Output & Ground-truth & Code-DKT-TC & TIK-TOC\\

\midrule

\multirow{-1.8}{*}{\input{student_codes/qual_codegen_helps_tcpred_gt_code}} & \multirow{-1.8}{*}{\input{student_codes/qual_codegen_helps_tcpred_tiktoc_code}} & `aaxbby' & `true'& \multicolumn{1}{c|}{\textcolor{OliveGreen}{\cmark}} & \multicolumn{1}{c|}{\textcolor{OliveGreen}{\cmark}} & \multicolumn{1}{c}{\textcolor{OliveGreen}{\cmark}}\\

&& `' & `true'& \multicolumn{1}{c|}{\textcolor{OliveGreen}{\cmark}} & \multicolumn{1}{c|}{\textcolor{CornellRed}{\xmark}} & \multicolumn{1}{c}{\textcolor{OliveGreen}{\cmark}}\\

&& `aaxbb' & `false'& \multicolumn{1}{c|}{\textcolor{OliveGreen}{\cmark}} & \multicolumn{1}{c|}{\textcolor{OliveGreen}{\cmark}} & \multicolumn{1}{c}{\textcolor{OliveGreen}{\cmark}}\\

&& `yaaxbb' & `false'& \multicolumn{1}{c|}{\textcolor{OliveGreen}{\cmark}} & \multicolumn{1}{c|}{\textcolor{OliveGreen}{\cmark}} & \multicolumn{1}{c}{\textcolor{OliveGreen}{\cmark}}\\

&& `yaaxbby' & `true'& \multicolumn{1}{c|}{\textcolor{OliveGreen}{\cmark}} & \multicolumn{1}{c|}{\textcolor{OliveGreen}{\cmark}} & \multicolumn{1}{c}{\textcolor{OliveGreen}{\cmark}}\\

&& `xaxxbby' & `true'& \multicolumn{1}{c|}{\textcolor{OliveGreen}{\cmark}} & \multicolumn{1}{c|}{\textcolor{OliveGreen}{\cmark}} & \multicolumn{1}{c}{\textcolor{OliveGreen}{\cmark}}\\

&& `xaxxbbyx' & `false'& \multicolumn{1}{c|}{\textcolor{OliveGreen}{\cmark}} & \multicolumn{1}{c|}{\textcolor{OliveGreen}{\cmark}} & \multicolumn{1}{c}{\textcolor{OliveGreen}{\cmark}}\\

&& `xxbxy' & `true'& \multicolumn{1}{c|}{\textcolor{OliveGreen}{\cmark}} & \multicolumn{1}{c|}{\textcolor{OliveGreen}{\cmark}} & \multicolumn{1}{c}{\textcolor{OliveGreen}{\cmark}}\\

&& `bbb' & `true'& \multicolumn{1}{c|}{\textcolor{OliveGreen}{\cmark}} & \multicolumn{1}{c|}{\textcolor{CornellRed}{\xmark}} & \multicolumn{1}{c}{\textcolor{OliveGreen}{\cmark}}\\

\bottomrule
\end{tabular}
\end{table}

\paragraph{Heatmap of test case pass/fail predictions.}
Table~\ref{tab:qualitative_heatmap} shows a heatmap of test case pass/fail probabilities predicted by TIKTOC, for a student attempting four problems, $P_{20}$, $P_{17}$, $P_{22}$, and $P_{21}$ in a row. These problems mainly cover the KC of if/else conditionals, requiring students to use multiple such conditionals correctly in their code. Problems $P_{20}$ and $P_{17}$ are similar, with $P_{17}$ being the easiest among the four problems; $P_{22}$ involves the additional KC of writing a helper method, and $P_{21}$ is the most difficult, requiring students to consider repetition and ordering of the input. Each cell in the heatmap represents a test case and is labeled with the actual pass/fail status. Cell colors depict the TIKTOC predicted probability of the student passing the test case, with darker colors corresponding to higher probabilities. Each column represents a subset of test cases associated with a problem since there are too many to visualize together. We also note that test cases differ across problems, which means cells in the same row may not perfectly align. Instead, we align them manually by grouping test cases across problems with similar difficulty or underlying KC. 

For problem $P_{20}$, TIKTOC correctly predicts the student will pass all easy test cases that check whether there is an else block that returns the sum of the three input integers. TIKTOC correctly predicts the student will fail a hard edge test case, $(121,121,121)$, with repeated input integers. We see that the predicted code (shown in the middle of Table~\ref{tab:qualitative_heatmap}) is semantically close to the ground-truth student code, both containing a similar logical error. TIKTOC correctly predicts the student will pass all test cases for $P_{17}$, the easiest problem, which tests KCs practiced before, in $P_{20}$.

The student's past knowledge from $P_{17}$, $P_{20}$, and other past problems, does not transfer to the new KC, writing a helper method, introduced in problem $P_{22}$. We see that their code fails all test cases, an outcome correctly predicted by TIKTOC. On problem $P_{21}$, the hardest problem, TIKTOC correctly predicts that the student will pass easy test cases covered before, but fail a hard edge case, ($13,13,13$), of a similar KC from $P_{20}$, where the inputs are all equal. This logical error is also reflected in the predicted student code (shown in the right of Table~\ref{tab:qualitative_heatmap}) in a semantically similar manner to the ground-truth student code. 
Across these problems, we see the student's estimated knowledge levels on similar KCs generally increase monotonically. 

\begin{table}[ht]
\caption{Qualitative case study showing test case pass/fail predictions help code generation. TIKTOC leverages its knowledge of which test cases are predicted to fail to predict code that is similar to the ground-truth. As a result, the predicted code shows the student's error in assuming the three input integers are sorted. Subset of test cases shown. Code indentation changed for brevity.}
\label{tab:qualitative_testcasepred_helps_codegen}
\small
\centering
\renewcommand{\arraystretch}{1.1}
\begin{tabular}{p{0.04\linewidth}|p{0.05\linewidth}p{0.06\linewidth}|p{0.20\linewidth}|p{0.26\linewidth}|p{0.20\linewidth}}

\toprule

\multicolumn{6}{L{16cm}}{Problem: Given three ints, a, b, and c, one of them is small, one is medium and one is large. Return true if the three values are evenly spaced, so the difference between small and medium is the same as the difference between medium and large.}\\

\midrule

\multirow{2}{0.05cm}{Test Case Input} & \multicolumn{2}{c|}{\textcolor{OliveGreen}{\cmark}/\textcolor{CornellRed}{\xmark} Preds} & \multicolumn{1}{c|}{\multirow{3}{*}{Ground-truth Student Code}} & \multicolumn{1}{c|}{\multirow{3}{*}{OKT Predicted Student Code}} & \multicolumn{1}{c}{\multirow{3}{*}{TIKTOC Student Code}} \\ 
\cmidrule{2-3}
& Ground-truth & TIKTOC &&&\\

\midrule

$2,4,6$ & \multicolumn{1}{c}{\textcolor{OliveGreen}{\cmark}} & \multicolumn{1}{c|}{\textcolor{OliveGreen}{\cmark}} & \multirow{-1.8}{*}{\input{student_codes/qual_tcpred_helps_codegen_gt_code}} & \multirow{-1.8}{*}{\input{student_codes/qual_tcpred_helps_codegen_okt_code}} & \multirow{-1.8}{*}{\input{student_codes/qual_tcpred_helps_codegen_tiktoc_code}}\\

$4,6,2$ & \multicolumn{1}{c}{\textcolor{CornellRed}{\xmark}} & \multicolumn{1}{c|}{\textcolor{CornellRed}{\xmark}} &&&\\

$4,6,3$ & \multicolumn{1}{c}{\textcolor{OliveGreen}{\cmark}} & \multicolumn{1}{c|}{\textcolor{OliveGreen}{\cmark}} &&&\\

$6,2,4$ & \multicolumn{1}{c}{\textcolor{CornellRed}{\xmark}} & \multicolumn{1}{c|}{\textcolor{CornellRed}{\xmark}} &&&\\

$6,2,8$ & \multicolumn{1}{c}{\textcolor{OliveGreen}{\cmark}} & \multicolumn{1}{c|}{\textcolor{OliveGreen}{\cmark}} &&&\\

$2,2,2$ & \multicolumn{1}{c}{\textcolor{OliveGreen}{\cmark}} & \multicolumn{1}{c|}{\textcolor{OliveGreen}{\cmark}} &&&\\

$2,2,3$ & \multicolumn{1}{c}{\textcolor{OliveGreen}{\cmark}} & \multicolumn{1}{c|}{\textcolor{OliveGreen}{\cmark}} &&&\\

\bottomrule
\end{tabular}
\end{table}

\begin{table}[ht]
\caption{Heatmap of test case pass/fail predictions by TIKTOC for consecutive student attempts on four problems involving multiple if/else conditionals. Ground-truth pass/fail labels are shown as $1$/$0$. TIKTOC accurately predicts test case pass/fail outcomes.}
\label{tab:qualitative_heatmap}
\small
\centering
\begin{tabular}{p{0.36\linewidth}|p{0.28\linewidth}|p{0.28\linewidth}}

\toprule

\multicolumn{3}{L{16cm}}{Problem $20$: Write a function in Java that implements the following logic: Given 3 int values, a, b, and c, return their sum. However, if one of the values is the same as another of the values, it does not count towards the sum.}\\

\midrule

\multicolumn{3}{L{16cm}}{Problem $21$: Write a function in Java that implements the following logic: Given 3 int values, a, b, and c, return their sum. However, if one of the values is 13 then it does not count towards the sum and values to its right do not count. So for example, if b is 13, then both b and c do not count.}\\

\midrule

TIKTOC Test Case \textcolor{OliveGreen}{\cmark}/\textcolor{CornellRed}{\xmark} Prediction Heatmap

& 

TIKTOC Pred Student Code: $P_{20}$

& 

TIKTOC Pred Student Code: $P_{21}$

\\ 
\addlinespace[-10pt]
\midrule

\includegraphics[width=6cm]{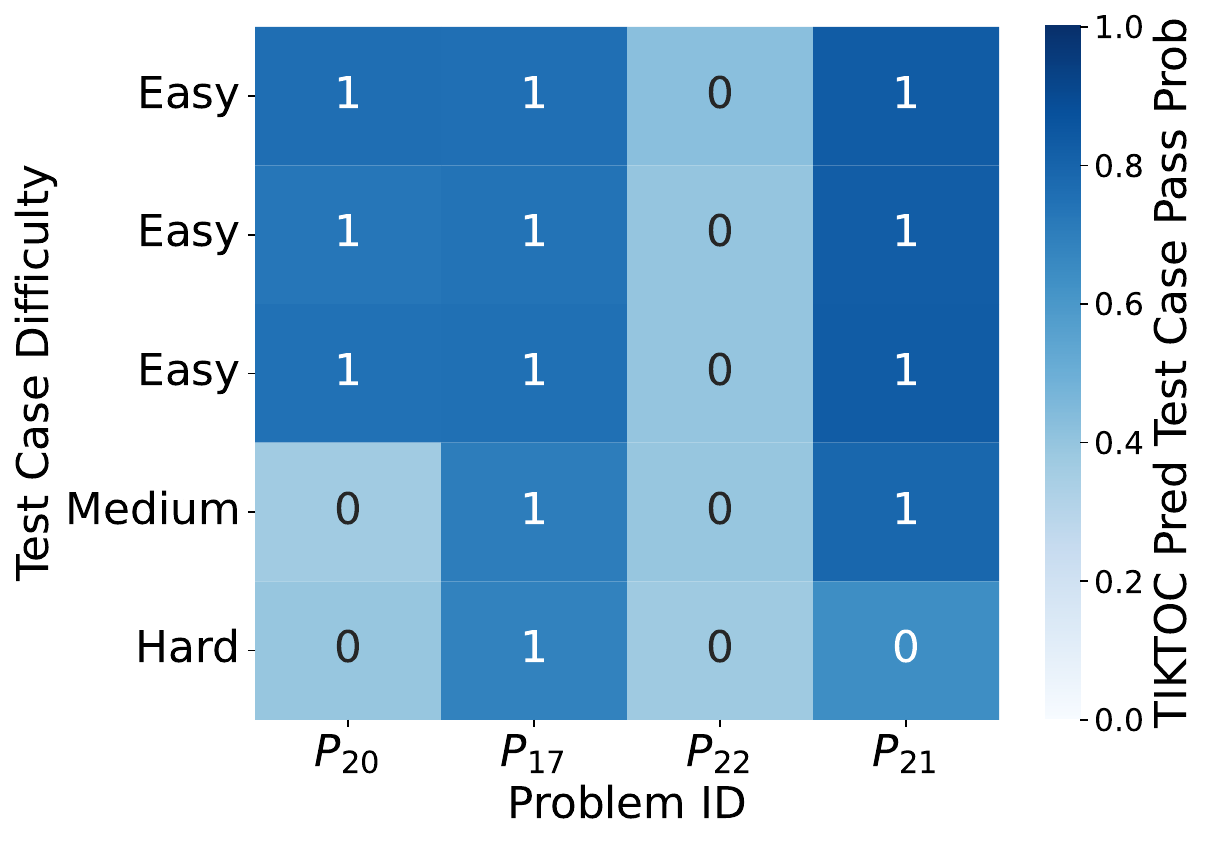} 
& 
\vspace{-3.5cm}
\input{student_codes/qual_heatmap_first_attempt_tiktoc_code}
& 
\vspace{-3.5cm}
\input{student_codes/qual_heatmap_last_attempt_tiktoc_code}
\\

\bottomrule
\end{tabular}
\end{table}

\subsection{Possible Use Cases in CS Education}

We now discuss how TIKTOC's test case pass/fail predictions of student code can be useful in real-world CS educational scenarios. For example, if a student is predicted to consistently fail test cases assessing operator precedence, this observation might suggest that they exhibit misconceptions of arithmetic expressions. Instructors can use TIKTOC's predictions to anticipate student errors on open-ended coding tasks through predicted test case pass/fail outcomes, even \textit{before} actually assigning these tasks to students. They can then prepare corresponding feedback, adjust the difficulty of the problems, and even design a suitable curriculum in advance. 

Writing good test cases for open-ended coding tasks can be challenging; the predictions from TIKTOC can perhaps help instructors on this task. For example, if a majority of students are predicted to pass all test cases for a problem, it may mean that existing test cases do not cover enough diversity among student codes. One can verify this postulate by checking clusters among the predicted code, following the visualization approach used in~\cite{okt}. In this case, instructors can explore creating new and harder test cases, possibly targeting edge cases, such as using an empty string as input, before verifying the predictions with TIKTOC again. For student support, if a student struggles to solve a problem, teachers can use TIKTOC to find students who also failed similar test cases but ultimately succeeded, and use their code trajectories to provide targeted, personalized hints. 

\section{Conclusions and Future Work}

In this paper, we proposed a challenging new KT task for open-ended coding tasks, which analyzes and predicts whether a student passes each test case in their code submission. To benchmark existing and new KT methods on this task, we have publicly released an augmented version of the real-world CodeWorkout dataset with test cases. We detailed TIKTOC, a novel KT method for this task, which uses a multi-task learning objective to jointly optimize 1) test case pass/fail prediction accuracy and 2) student code prediction accuracy. Through extensive experiments, we showed that TIKTOC outperforms existing state-of-the-art KT methods adapted to this task. Through qualitative case studies, we showed that our multi-task learning setup is effective and discussed potential use cases in real-world CS educational scenarios.

To the best of our knowledge, TIKTOC is among the first attempts to leverage test cases to model student learning open-ended coding tasks. We identify several limitations and avenues for future work. 
First, we can explore ways to effectively leverage the exact content of test cases to improve pass/fail prediction accuracy. 
Second, we can explore providing the ground-truth test case pass/fail labels of previous attempts by a student to the LSTM to estimate their knowledge state like DKT~\cite{dkt}.
Third, test cases can be labeled with knowledge concepts (KCs), with a training objective encouraging test case/KC learning curves (student error rates) to follow the power law of practice~\cite{shi2023kc, snoddy1926learning}. 
Fourth, TIKTOC does not explicitly control for fairness across students from different demographic groups and could incorporate fairness regularization into the training objective~\cite{Zafar:Fairness:2017}.

\section{Acknowledgements}
This work is partially supported by the NSF under grants 2215193, 2237676, and 2418657.

\bibliographystyle{plain}
\bibliography{main}

\end{document}

%% file: student_codes/qual_example_student_code.tex
\begin{lstlisting}
public String getSandwich(String str){
    int length = str.length();
    if(str.startsWith("bread")&&str.endsWith("bread"))
    {
        return str.substring(5, str.length() - 5);
    }
    else
    {
        return "";
    }
}\end{lstlisting}

%% file: student_codes/qual_codegen_helps_tcpred_gt_code.tex
\begin{lstlisting}
public boolean xyBalance(String str){
    int length=str.length()-1;
    char character;
    for(int i=length;i>=0;i--){
        character=str.charAt(i);
        if(character == 'x'){
            return false;
        }
        else if(character=='y'){
            return true;
        }
    }
    return true;  
}\end{lstlisting}

%% file: student_codes/qual_codegen_helps_tcpred_tiktoc_code.tex
\begin{lstlisting}
public boolean xyBalance(String str){
    int len=str.length()-1;
        char ch;
        for(int i=len; i>=0; i--){
            ch = str.charAt(i);
            if(ch == 'x')
                return false;
            else if(ch == 'y')
                return true;
        }
        return true;  
}\end{lstlisting}

%% file: student_codes/qual_tcpred_helps_codegen_gt_code.tex
\begin{lstlisting}
public boolean evenlySpaced
    (int a, int b, int c)
{
    if((a-b)==(b-c))
        return true;
    else
        return false;
}\end{lstlisting}

%% file: student_codes/qual_tcpred_helps_codegen_okt_code.tex
\begin{lstlisting}
public boolean evenlySpaced
    (int a, int b, int c){
    if(Math.abs(a-b)==Math.abs(b-c) 
    ||Math.abs(b-a)==Math.abs(a-c) 
    ||Math.abs(a-c)==Math.abs(c-b)){
        return true;
    }
    return false;
}\end{lstlisting}

%% file: student_codes/qual_tcpred_helps_codegen_tiktoc_code.tex
\begin{lstlisting}
public boolean evenlySpaced
(int a, int b, int c){
    int dif;
    dif = c-b;
    if( dif == b-a){
        return true;
    }
    else{
        return false;
    }
}\end{lstlisting}

%% file: student_codes/qual_heatmap_first_attempt_tiktoc_code.tex
\begin{lstlisting}
public int loneSum
    (int a, int b, int c) {
    if (a == b) {
        return c;
    }
    else if (a == c) {
        return b;
    }
    else if (b == c) {
        return a;
    }
    else {
        return a + b + c;
    }
}\end{lstlisting}

%% file: student_codes/qual_heatmap_last_attempt_tiktoc_code.tex
\begin{lstlisting}
public int luckySum
    (int a, int b, int c) {
    if (a == 13) {
        return c;
    }
    else if (b == 13) {
        return a;
    }
    else if (c == 13) {
        return a + b;
    }
    else {
        return a + b + c;
    }
}\end{lstlisting}